\documentclass[fleqn,twoside]{article}
\usepackage{espcrc2}
\usepackage{graphicx}

\title{Strange matrix elements of the nucleon}

\author{Randy Lewis\address{Department of Physics, University of Regina,
                            Regina, SK, S4S 0A2, Canada},
        W. Wilcox\address{Department of Physics, Baylor University, Waco, TX,
                          76798-7316, U.S.A.}
        and
        R. M. Woloshyn\address{TRIUMF, 4004 Wesbrook Mall, Vancouver, BC,
                               V6T 2A3, Canada}}
       
\begin{document}

\begin{abstract}
Results for the disconnected contributions to matrix elements of the
vector current and scalar density have been obtained for the nucleon
from the Wilson action at $\beta=6$ using a stochastic estimator
technique and 2000 quenched configurations.  Various methods for analysis
are employed and chiral extrapolations are discussed.
\vspace{1pc}
\end{abstract}

\maketitle

\section{MOTIVATION}

Prediction, measurement and understanding of strangeness in the nucleon
have proven to be significant challenges for both theory and experiment.
Of particular interest at present are the strangeness electric and magnetic
form factors.  Experimental results have been reported
by SAMPLE\cite{SAMPLE} and HAPPEX\cite{HAPPEX}, 
and other groups are also planning experiments.\cite{A4G0}

A number of lattice QCD studies have been reported during the past few
years\cite{others,emi}, though the conclusions are somewhat varied.
In the present work we report the results of a high statistics simulation.
Different analysis techniques are studied and chiral
extrapolations are discussed.  The strangeness scalar density and
electric and magnetic form factors are all analyzed together,
and results are compared to experimental data.

\section{SIMULATIONS AND ANALYSIS}

The nucleon's strangeness form factors arise from a disconnected strange
quark loop that we compute stochastically using real $Z_2$ noise\cite{DongLiu}.
To reduce the variance, the first four terms of the perturbative quark
matrix are subtracted.\cite{subtraction}  (For the scalar density, the first
five terms are subtracted.)  The three matrix elements of interest are
\begin{equation}
M_{\{S,M,E\}}(t,\vec{q}) =
\left\{G_S^{(s)},\frac{q_jG_M^{(s)}}{E_q+m},G_E^{(s)}\right\}
\end{equation}
where $\left<N\left|\bar{s}s\right|N\right>=Z_SG_S^{(s)}$,
$Z_S\approx 1-3\kappa_v/(4\kappa_c)$,
and the $M_X(t,\vec{q})$ can be extracted from the ratios
\begin{equation}
R_X(t,t^\prime,\vec{q}) =
     \frac{G_X^{(3)}(t,t^\prime,\vec{q})G^{(2)}(t^\prime,\vec0)}
          {G^{(2)}(t,\vec0)G^{(2)}(t^\prime,\vec{q})}
\end{equation}
of 2 and 3-point correlators by various methods:
\begin{equation}\label{differential}
\sum_{t^\prime=1}^{t+1}\left[R_X(t,t^\prime,\vec{q})
-R_X(t\!-\!1,t^\prime,\vec{q})\right] \to M_X(t,\vec{q}),
\end{equation}
\begin{equation}
\sum_{t^\prime=1}^{t}R_X(t,t^\prime,\vec{q}) \to {\rm constant}
 + tM_X(t,\vec{q}),
\end{equation}
\begin{equation}
\sum_{t^\prime=1}^{t_{fixed}}R_X(t,t^\prime,\vec{q}) \to {\rm constant}
 + tM_X(t,\vec{q}).
\end{equation}
Any authentic signal should be visible with each of these methods, and
should be consistent among all of them.\cite{emi}

\begin{figure}[thb]
\includegraphics*[width=75mm]{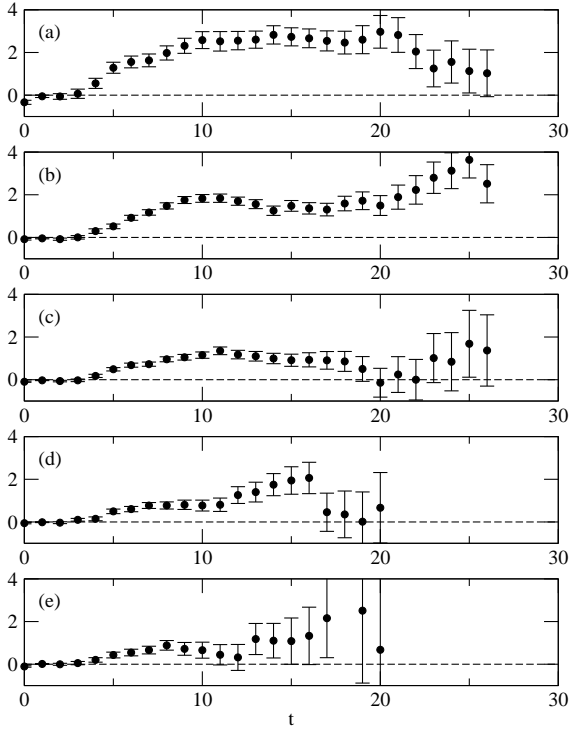}
\caption{$M_S(t,\vec{q})$ from Eq.~(\protect\ref{differential}) with
         $\kappa_v=\kappa_l=0.152$.  From top to
         bottom, the plots show $\vec{q}_L^{\,2}=0$, 1, 2, 3
         and 4, where $\vec{q}_L\equiv(10a/\pi)\vec{q}$.}\label{figscalar}
\vspace{-2mm}
\end{figure}
Simulations have been performed on $20^3\times32$ lattices using the
Wilson action with $\beta=6$.  Dirichlet time boundaries are used for quarks
with the source five timesteps from the lattice boundary.
Valence quarks use $\kappa_v=0.152$, 0.153 and 0.154.  The chiral limit
is $\kappa_c\approx0.1571$ and our $\kappa_v$ values are in the strange region.
The mass of the quark in 
the loop will be held fixed at a value corresponding to $\kappa_l=0.152$
and the analysis will be based on 2000 configurations with 60 $Z_2$ noises
per configuration.  Consistent results (not shown in this brief article)
have been obtained from the
analysis of 100 configurations with $\kappa_l=0.154$ and
200 $Z_2$ noises per configuration.

Results for $M_S(t,\vec{q})$ from Eq.~(\ref{differential}) are shown in
Fig.~\ref{figscalar}.  For each $\vec{q}$ considered, a clear plateau begins
about 10 timesteps from the source.  (The source is at $t=0$ in the plots.)
On the other hand, 
results for $M_M(t,\vec{q})$ and $M_E(t,\vec{q})$ are consistent with zero
for each $\vec{q}$ and $\kappa_v$ as shown in Table~\ref{tableraw}.

\section{CHIRAL FITS}

\begin{figure}[thb]
\includegraphics*[width=75mm]{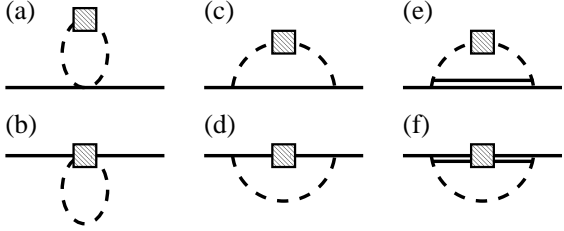}
\caption{Leading loop diagrams from quenched chiral perturbation theory.
         Dashed, solid and double lines denote octet mesons, octet baryons
         and decuplet baryons respectively.  A shaded box denotes a current
         insertion.}\label{figloops}
\vspace{-5mm}
\end{figure}

Quark mass and momentum dependences can be calculated 
analytically within quenched chiral perturbation theory
from the diagrams of Fig.~\ref{figloops}
as well as all tree-level counterterms.

\begin{figure}[thb]
\includegraphics*[width=75mm]{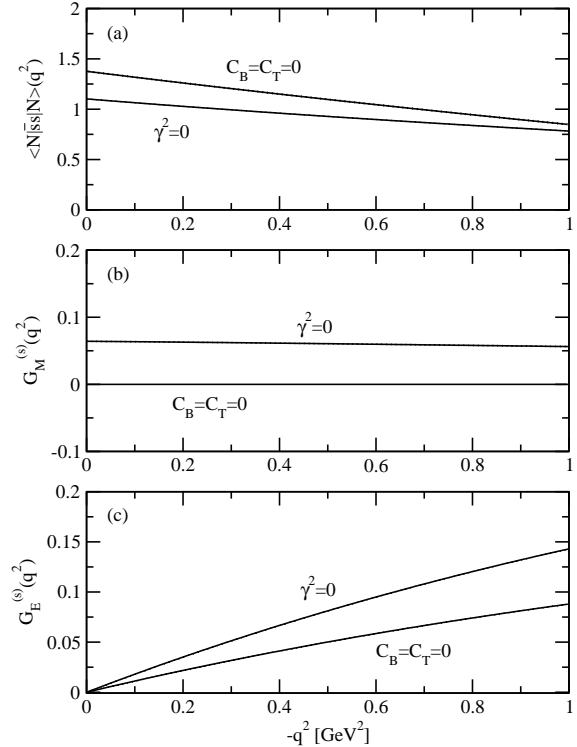}
\caption{Lattice results extrapolated to the physical hadron
         masses. \vspace{-2mm}}\label{figfinal}
\end{figure}

\begin{table}[thb]
\caption{Fits to the matrix elements of Eq.~(\protect\ref{differential})
         beginning 10 timesteps from the source.
         $\vec{q}_L \equiv (10a/\pi)\vec{q}$.}\label{tableraw}
\begin{tabular}{ccr@{.}lr@{.}lr@{.}l}
\hline\hline
$\kappa_v$ & $\vec{q}_L^{\,2}$ & \multicolumn{6}{c}{$\kappa_l=0.152$} \\
\cline{3-8}
 & & \multicolumn{2}{c}{$G_S^{(s)}$} & \multicolumn{2}{c}{$G_M^{(s)}$}
   & \multicolumn{2}{c}{$G_E^{(s)}$} \\
\hline
\hline
0.152 & 0 & 2&6(4)  & \multicolumn{2}{c}{---} & -0&009(13) \\
      & 1 & 1&7(2)  &  0&007(16)              & -0&008(8) \\
      & 2 & 1&2(2)  & -0&018(14)              &  0&012(10) \\
      & 3 & 1&1(5)  & -0&014(23)              &  0&008(17) \\
      & 4 & 0&7(6)  &  0&004(31)              &  0&026(40) \\
\hline
0.153 & 0 & 2&7(5)  & \multicolumn{2}{c}{---} & -0&010(15) \\
      & 1 & 1&8(3)  &  0&012(22)              & -0&011(10) \\
      & 2 & 1&3(2)  & -0&021(20)              &  0&015(14) \\
      & 3 & 1&2(6)  & -0&018(32)              &  0&008(22) \\
      & 4 & 0&7(8)  &  0&005(48)              &  0&029(56) \\
\hline
0.154 & 0 & 2&9(5)  & \multicolumn{2}{c}{---} & -0&013(19) \\
      & 1 & 1&8(3)  &  0&019(33)              & -0&014(15) \\
      & 2 & 1&3(3)  & -0&022(31)              &  0&019(21) \\
      & 3 & 1&5(9)  & -0&029(53)              &  0&008(32) \\
      & 4 & 0&8(11) &  0&010(82)              &  0&021(81) \\
\hline
\hline
\end{tabular}
\end{table}
There are six independent parameters and we consider two
extreme limits: either the quenched $\eta^\prime$ is absent
($\gamma=0$), or $\gamma\neq0$ but the
non-$\eta^\prime$ parameters are set to zero in the loops.
The physical situation must lie between
these two extremes, and so we expect the true values
to be somewhere between the curves in
Fig.~\ref{figfinal}.  (For another recent discussion of chiral fitting,
see Ref.~\cite{CS}.)

\section{COMPARING TO EXPERIMENT}

The results of Figure~\ref{figfinal} compare to the existing experimental
results as follows:
\begin{equation}
G_M^{(s)}(q_1^2) = \left\{\begin{array}{ll}
 0.14 \pm 0.29 \pm 0.31, & {\rm ~Ref.~\cite{SAMPLE}} \\
 0.03 \pm 0.03,          & {\rm ~Fig.~\ref{figfinal}} \end{array}\right.
\end{equation}
\begin{eqnarray}
G_E^{(s)}(q_2^2) + 0.39G_M^{(s)}(q_2^2)
\hspace{-34mm}
\nonumber \\
&&= \left\{\begin{array}{ll}
 0.025 \pm 0.020 \pm 0.014, & {\rm ~Ref.~\cite{HAPPEX}} \\
 0.027 \pm 0.016,          & {\rm ~Fig.~\ref{figfinal}} \end{array}\right.
\end{eqnarray}
where $-q_1^2=0.1$ GeV$^2$ and $-q_2^2=0.477$ GeV$^2$.
For the scalar density, we find
\begin{equation}
(m_s/m_N)\left<N\left|\bar{s}s\right|N\right>(0) = 0.15(2),
\end{equation}
which can be compared to the result of 0.195(9) reported in Ref.~\cite{DLL}.

We have not attempted to estimate the sizes of systematic
uncertainties in our results due, for example, to quenching and to
the use of chiral extrapolations for valence quarks in the strange region.
The raw lattice data of
Table~\ref{tableraw} show, at best, only tiny strange quark effects
over the range of momenta and quark masses that were
studied. Large strange-quark loop contributions in the vector current
matrix elements should be considered unlikely.

\section*{ACKNOWLEDGEMENTS}

This work was supported in part by the National Science Foundation under
grant 0070836, the Baylor Sabbatical Program, and the Natural Sciences and
Engineering
Research Council of Canada.  Some of the computing was done on hardware funded
by the Canada Foundation for Innovation with contributions from Compaq Canada,
Avnet Enterprise Solutions and the Government of Saskatchewan.


\begin{thebibliography}{9}
\bibitem{SAMPLE} R. Hasty et. al., Science 290, 2117 (2000).
\bibitem{HAPPEX} K.A. Aniol et. al., Phys. Lett. B509, 211 (2001).
\bibitem{A4G0} For example, the A4 Collaboration at MAMI and the
               G0 Collaboration at Jefferson Lab.
\bibitem{others} S.J. Dong, K.F. Liu and A.G. Williams, Phys. Rev. D58,
                 074504 (1998);
                 D.B. Leinweber and A.W. Thomas, Phys. Rev. D62, 07505 (2000);
                 N. Mathur and S.J. Dong, Nucl. Phys. (P.S.) 94, 311
                 (2001);
                 W. Wilcox, Nucl. Phys. (P.S.) 94, 319 (2001).
\bibitem{emi} R. Lewis, W. Wilcox and R.M. Woloshyn, hep-ph/0201190 (2002).
\bibitem{DongLiu} S.J. Dong and K.F. Liu, Nucl. Phys. (P.S.) 26,
                  353 (1992); Phys. Lett. B328, 130 (1994).
\bibitem{subtraction} C. Thron, S.J. Dong, K.F. Liu and H.P. Ying, Phys.
                      Rev. D57, 1642 (1998);
                      W. Wilcox, hep-lat/9911013, in {\it Numerical Challenges
                      in Lattice Quantum Chromodynamics}, edited by A. Frommer
                      et. al. (Springer Verlag, Heidelberg, 2000);
                      Nucl. Phys. (P.S.) 83, 834 (2000);
                      C. Micheal, M.S. Foster and C. McNeile, Nucl. Phys.
                      (P.S.) 83, 185 (2000).
\bibitem{CS} J.W.~Chen and M.J.~Savage, hep-lat/0207022.
\bibitem{DLL} S.J. Dong, J.F. Laga\"e and K.F. Liu, Phys. Rev. D54, 5496
              (1996).
\end{thebibliography}
\end{document}